\documentclass[pra,twocolumn,aps,superscriptaddress,showpacs,epsfig]{revtex4}
\usepackage{epsfig,graphicx}
\usepackage{amsfonts}
\usepackage{amssymb}
\usepackage{amstext}
\usepackage{amsmath}

\def\be{\begin{equation}}
\def\ee{\end{equation}}
\def\ba{\begin{eqnarray}}
\def\ea{\end{eqnarray}}
\def\bw{\begin{widetext}}
\def\ew{\end{widetext}}

\begin{document}

\title{Formation and Transformation of Vector-Solitons in Two-Species Bose-Einstein Condensates with
Tunable Interaction
}

\author{Xunxu Liu}
\affiliation{Beijing National Laboratory for Condensed Matter
Physics, Institute of Physics, Chinese Academy of Sciences, Beijing
100080, China}
\affiliation{Department of Physics and Centre of
Computational Science and Engineering, National University of
Singapore, 117542, Singapore}
\author{Han Pu}
\affiliation{Department of Physics and Astronomy, and Rice Quantum Institute, Rice
University, Houston, Texas 77251-1892, USA}
\author{Bo Xiong}
\affiliation{Department of Physics and Centre of Computational Science and Engineering,
National University of Singapore, 117542, Singapore}
\author{W. M. Liu}
\affiliation{Beijing National Laboratory for Condensed Matter Physics, Institute of
Physics, Chinese Academy of Sciences, Beijing 100080, China}
\author{Jiangbin Gong}
\affiliation{Department of Physics and Centre of Computational Science and Engineering,
National University of Singapore, 117542, Singapore}

\date{\today }

\begin{abstract}
Under a unified theory we investigate the formation of various types
of vector-solitons in two-species Bose-Einstein condensates with
arbitrary scattering lengths.
%We show that, for quasi-one-dimension homogeneous systems,
%exact solutions exist even in the regime where the nonlinear system
%is non-integrable.
We then show that by tuning the interaction parameter via Feshbach
resonance, transformation between different types of vector solitons
is possible.  Our results open up new ways in the quantum control of
multi-species Bose-Einstein condensates.
\end{abstract}

 \maketitle

Since the first realization of Bose-Einstein condensate (BEC), a
tremendous amount of research has taken place in this
interdisciplinary field \cite{first_experiment1,book}. One specific
topic of wide interest is multi-component BECs, which possess
complicated quantum phases \cite{AlonPRL} and properties not seen in
individual components. Theoretical \cite{hoshenoy,esry,PuHPRL881130}
and experimental \cite{cornell,ModugnoScience294,hall} studies  have
shown that inter-species interaction plays crucial roles in these
systems. Recently, two-species BECs with tunable interaction
\cite{Italy_twospecies,Italy_twospecies1} were successfully produced
. This important progress motivates further explorations of
two-species BECs.

Solitons represent a highly nonlinear wave phenomenon with unique
propagation features and have attracted great interests from many
different fields, e.g., nonlinear optics \cite{opt}.
%One important
%aspect of BEC dynamics is matter-wave solitons described by
%nonlinear Schr\"{o}dinger equations
Because the nonlinear Schr\"{o}dinger equation governing matter-wave
solitons \cite{book,darkexp,Science2961290,nature} is similar to
those for solitons in other contexts, well-established
understandings of solitons can be applied to atomic BECs to achieve
better control of matter waves. However, BEC systems possess many
unique features of their own and make it possible to study soliton
phenomena under circumstances that cannot be easily realized in
other fields.

In this paper, we provide a unified theory to investigate various
types of vector solitons in two-species condensates with arbitrary
scattering lengths. Remarkably, we find that by tuning the
inter-species interaction via the well-established Feshbach
resonance technique \cite{Italy_twospecies}, different types of
vector-solitons can be realized and even transformed to each other.
In addition, these vector solitons can be dynamically stable in the
presence of soft trapping potentials. Our theoretical work opens up
new possibilities in the quantum control of multi-species BECs and
other related complex systems.

The mean-field dynamics of a two-species BEC is
governed by the following equations \cite{hoshenoy,esry,PuHPRL881130,book}:
\begin{eqnarray*}
&&i\hbar \frac{\partial \psi _{1}}{\partial t}=[-\frac{\hbar^{2}\nabla ^{2}}{2m_{1}}%
+V_{1}+U_{11}|\psi _{1}|^{2}+U_{12}|\psi _{2}|^{2}]\psi _{1},
\notag \\
&&i\hbar \frac{\partial \psi _{2}}{\partial t}=[-\frac{\hbar ^{2}\nabla ^{2}}{2m_{2}}%
+V_{2}+U_{21}|\psi _{1}|^{2}+U_{22}|\psi _{2}|^{2}]\psi_{2},
%\label{GP}
\end{eqnarray*}%
where the condensate wave functions are normalized by particle
numbers $N_i\equiv \int  |\psi_i|^2 d^{3}{\bf r}$, $U_{ii}=4\pi
\hbar ^{2}a_{ii}/m_{i}$ and  $U_{12}=U_{21} =2\pi \hbar
^{2}a_{12}/m$ represent intra- and inter-species interaction
strengths respectively, with $a_{ij}$ being the corresponding
scattering lengths and $m$ the reduced mass. The trapping
potentials are assumed to be $V_{i}=m_{i}[\omega
_{ix}^{2}x^{2}+\omega _{i\perp}^{2}(y^{2}+z^{2})]/2$. Further
assuming $\omega_{i\perp} \gg \omega_{ix}$ such that the transverse
motion of the condensates are frozen to the ground state of the
transverse harmonic trapping potential, the system becomes
quasi-one-dimensional. Integrating out the transverse coordinates,
the resulting equations for the axial wave functions
$\tilde{\psi}_{1,2}(x)$ in dimensionless form can be written as
\begin{eqnarray}
i\frac{\partial \tilde{\psi}_{1}}{\partial t} \!\!&=&\!\!
[-\frac{1}{2}\frac{\partial^{2}}{\partial x^{2}}+\frac{\lambda _{1}^{2}}{2} x^{2}
+b_{11}|\tilde{\psi}_{1}|^{2}+b_{12}|\tilde{\psi}_{2}|^{2} ] \tilde{\psi}_{1} \label{DGP1},  \\
i\frac{\partial \tilde{\psi}_{2}}{\partial t} \!\!&=&\!\! [-\frac{\kappa }{2}\frac{%
\partial ^{2}}{\partial x^{2}}+\frac{\lambda _{2}^{2}}{2\kappa }
x^{2}+b_{21}|\tilde{\psi}_{1}|^{2}+b_{22}|\tilde{\psi}_{2}|^{2} ]
\tilde{\psi}_{2},  \label{DGP2}
\end{eqnarray}
where we have chosen $\sqrt{\hbar/(m_1 \omega_{1\perp})}$ and
$2\pi/\omega_{1\perp}$ to be the units for length and time,
respectively; and $\tilde{\psi}_{1,2}$ is normalized such that $%
\int |\tilde{\psi}_{1}|^{2}dx=1$ and $\int
|\tilde{\psi}_{2}|^{2}dx=N_2/N_1$. Other parameters in Eqs.~
(\ref{DGP1}) and (\ref{DGP2}) are defined as: $b_{11}=2a_{11}N_{1}$,
$b_{12}=b_{21}=2m_{1}a_{12}N_{1}/[(1+\gamma )m]$,
$b_{22}=2a_{22}m_{1}N_{1}\gamma /m_{2}$,  $\gamma =\omega
_{2\perp}^{{}}/\omega _{1\perp}^{{}}$, $\lambda _{1}=\omega
_{1x}/\omega _{1\perp}$, $\lambda _{2}=\omega _{2x}/\omega
_{1\perp}$, and $\kappa =m_{1}/m_{2}$.

%Typically, in $^{7}$Li, N=5000, a$_{s}=-0.21nm$, the time unit is 0.224ms,
%length unit is 1.42$\mu m$\cite{Science2961290}.

{\em A unified treatment of different vector-soliton solutions {\rm
--}} When the longitudinal trapping potential is neglected
($\lambda_i=0$), Eqs.~(\ref{DGP1}) and (\ref{DGP2}) are perfectly
integrable under the condition $\kappa=1$ (i.e., $m_1=m_2$) and $
b_{11}=b_{12}=b_{21}=b_{22}$ \cite{int}, allowing for a general
procedure to construct two-component vector-soliton solutions
in the
form of
``dark-dark" \cite{PRE554773}, ``bright-dark" \cite%
{PRL87010401} and ``bright-bright" solitons \cite{PRA48599}. When
integrability is destroyed, previous studies show that distorted
versions of soliton solutions exist, but closed form can only be
given for special cases and often for one particular type of vector soliton \cite{special}.
Here we show
that, even when the above-mentioned integrability condition is
violated (e.g., for arbitrary interaction strengths $b_{ij}$ and mass ratio $\kappa$), there still exists in general a specific class of exact
solutions that include all types of vector solitons,
so
long as $b_{12}^2 \neq b_{11}b_{22}$. The type of vector soliton is
determined by the values of $b_{ij}$ and $\kappa$.

The vector-soliton solution can be found by inserting an appropriate
ansatz (see below) into Eqs.~ (\ref{DGP1}) and (\ref{DGP2}).
Defining the following two quantities:
\begin{eqnarray*}
C_1 & \equiv & (b_{22}-\kappa b_{12} )/(b^2_{12}-b_{11}b_{22})
\,,\\
C_2 & \equiv & (b_{12}- \kappa b_{11})/(b^2_{12}-b_{11}b_{22}) \,,
\end{eqnarray*}
the conditions under which various vector-soliton solutions exist
are found to be:
\begin{eqnarray*}
&&C_1 >0, \,C_2<0:\;\;{\rm bright-bright}\;({\rm BB}), \\
&&C_1>0,\,C_2>0:\;\;{\rm bright-dark}\;({\rm BD}), \\
&&C_1<0, \,C_2>0:\;\;{\rm dark-dark}\;({\rm DD}), \\
&&C_1<0,\,C_2<0:\;\;{\rm dark-bright}\;({\rm DB}).
\end{eqnarray*}
Explicit expressions for these vector-solitons are then found from
the ansatz we use.  In particular, the BB solution (i.e., bright
soliton for species 1 and bright soliton for species 2; analogous
convention applies to all other solutions) is
 given by
\begin{eqnarray*}
\tilde{\psi}_{1B}\!\!  &=\!\!&\eta\sqrt{C_1} \, {\rm sech}(\eta
x-\eta vt)\,
e^{i\left[vx+\left({\eta^{2}}-{%
v^{2}}\right)t/2\right]}\,,   \\
\tilde{\psi}_{2B}\!\!  &=\!\! & \eta \sqrt{-C_2} \,{\rm sech}(\eta
x-\eta vt)\, e^{i \left[{v}x/{\kappa }+\left({\kappa
}\eta^{2}-v^{2}/\kappa\right)t/2\right]}.
\end{eqnarray*}
The BD solution is given by
\begin{subequations}
\label{BD}
\begin{eqnarray}
\tilde{\psi}_{1B}\!\!  &=\!\! &\eta \sqrt{C_1}  \, {\rm sech}(\eta
x-\eta vt)\,
e^{i\left[vx+f_1t \right]}\,,  \label{B1} \\
\tilde{\psi}_{2D} \!\! &=&\!\! [ iv\sqrt{C_2}/\kappa+\eta \sqrt{C_2}
\,\tanh(\eta x\!\!-\!\!\eta vt) ] e^{i f_2 t }, \label{D2}
\end{eqnarray}
\end{subequations}
with $f_1=\left({\eta^{2}}-{%
v^{2}}\right)/2 -b_{12}C_2(\eta^2+v^2/\kappa^2)$ and $f_2 =-b_{22}
C_2(\eta^2+v^2/\kappa^2)$. The DD solution is found to be
\begin{eqnarray*}
\tilde{\psi}_{1D} \!\! &=&\!\! [iv\sqrt{-C_1}+\eta \sqrt{-C_1}
\,\tanh(\eta x-\eta vt) ]\,
e^{if_1t }\,,   \\
\tilde{\psi}_{2D} \!\! &=&\!\! [ iv\sqrt{C_2}/\kappa+\eta \sqrt{C_2}
\,\tanh(\eta x-\eta vt) ]\, e^{i f_2 t }\,,
\end{eqnarray*}
with $f_1 = -\eta^2-v^2(-b_{11}C_1+b_{12}C_2/\mu^2)$ and $f_2 =
-\mu\eta^2-v^2(-b_{21}C_1+b_{22}C_2/\mu^2)$. Finally, the DB
solution can be obtained from the BD solution (\ref{BD}) by
exchanging indices 1 and 2, and let $C_{1,2} \rightarrow -C_{1,2}$.
In all these cases, the parameter $\eta$ determines the width of the
soliton and can be found by the normalization condition for
$\tilde{\psi}_{1,2}$. The parameter $v$ gives the velocity of the
soliton. We note that the BB solution under the condition $b_{11},
b_{22}>0$ and $b_{12}<0$, but not the other cases, has been studied
by P\'{e}rez-Garc\'{i}a and Beitia \cite{pra}. The exact expressions
for vector solitons found here bear an apparent resemblance to those
for integrable systems \cite{int}.

\begin{figure}[t]
\includegraphics[width=8cm]{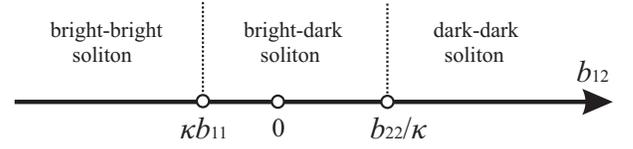}
\caption{Phase diagram of vector solitons vs. inter-species
interaction $b_{12}$ for the system with $b_{11}<0$ and $b_{22}>0$.
As described in the text, the unit of $b_{ij}$ is $\sqrt{\hbar/(m_1
\omega_{1\perp})}$. In the examples shown later in the paper, the
two species experience the same trapping frequency:
$\omega_{1\perp}=\omega_{2\perp}=2\pi \times 710$Hz. The number of
atoms in the first species ($^7$Li) is assumed to be 5000, which
results in the dimensionless intra-species interaction strength
$b_{11}=-1.47$ and $b_{22}=5.89$. The two critical values of the
inter-species interaction strength are therefore $\kappa
b_{11}=-0.45$ and $b_{22}/\kappa = 19.35$.} \label{demo-region}
\end{figure}

{\em Dynamics and dynamical stability {\rm --}} To be specific, we
take $^7$Li as species 1 and $^{23}$Na as species 2, with fixed and
realistic intra-species interaction strength $b_{11}<0$, $b_{22}>0$
and variable inter-species interaction strength $b_{12}$. Figure
~\ref{demo-region} depicts the ``phase diagram'' for different
regimes of vector solitons as $b_{12}$ is varied. As is indicated in
Fig. ~\ref{demo-region}, the system supports BB ($b_{12}<\kappa
b_{11}$), BD ($\kappa b_{11}<b_{12}<b_{22}/\kappa$) and DD
($b_{12}>b_{22}/\kappa$) vector soliton, while the condition for DB
soliton can never be satisfied for this particular $^7$Li--$^{23}$Na
system.

To study the dynamical stability of the vector-soliton solutions
given above, we first calculate the collective excitation
frequencies by solving the corresponding Bogoliubov-de Gennes
equation \cite{PuHPRL881130}. Our calculations show that, the
excitation frequencies of the BB and BD vector-solitons are all
real, while those of the DD vector-soliton contain complex values.
%Since the presence of complex excitation frequencies is an
%indication that the system is dynamically unstable,
We hence conclude that, for the $^7$Li-$^{23}$Na system, the BB and
BD vector-solitons are dynamically stable, while the DD
vector-soliton is dynamically unstable.

\begin{figure}[t]
\includegraphics[width=8.4cm]{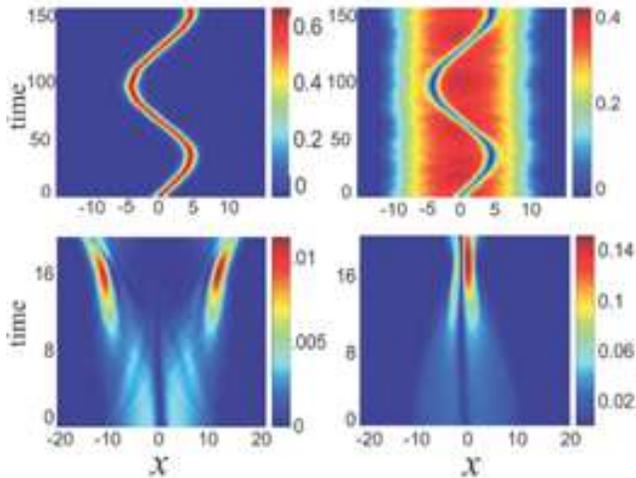}
\caption{(Color online) Dynamics of vector solitons inside a weak
longitudinal trap with $\protect\lambda_1=\protect\lambda_2=0.1$.
Other parameters are specified in Fig.~\ref{demo-region}. Units for
$x$ and time are $\sqrt{\hbar/(m_1 \omega_{1\perp})}$ and
$2\pi/\omega_{1\perp}$. The left and right plots represent the
longitudinal density profiles of the first and second species,
respectively. Upper panel: evolution of a BD vector soliton with
$b_{12}=2$ initially located at $x=0$ and velocity $v=0.1$. Lower
panel: evolution of a DD vector soliton with $b_{12}=30$. For the
system under study, the BD (DD) vector soliton is dynamically stable
(unstable).} \label{BDDD}
\end{figure}

This stability analysis is confirmed by direct numerical simulations
of Eqs.~(\ref{DGP1}) and (\ref{DGP2}). Computational examples are
shown in Fig.~\ref{BDDD}. In our simulation, a harmonic trapping
potential is added along the longitudinal direction to represent a more realistic situation. %The width of the soft trapping potential is
%chosen to be much wider than the width of the solitons.
The trapping potential is weak such that its variation across the
soliton scale is negligible. Such a soft trapping potential
spatially confines the solitons without affecting their essential
properties. The upper panel of Fig.~\ref{BDDD} shows the evolution
of a BD vector soliton. For the initial condition, we use the exact
solution of Eq.~(\ref{B1}) for the bright component, while in order
to confine the dark soliton inside the trap, we multiply
Eq.~(\ref{D2}) by a Thomas-Fermi profile with an inverted parabolic
shape to simulate the dark component which is a quite standard
practice \cite{PRL87010401}. The presence of the weak longitudinal
trap causes the vector soliton to oscillate while maintaining the
overall shape as shown in the upper panel of Fig.~\ref{BDDD}. By
contrast, the lower panel of Fig.~\ref{BDDD} shows that the DD
soliton is dynamically unstable
--- an initial DD vector-soliton is quickly destroyed and the two species tend
to phase separate into different spatial regions due to the large
inter-species repulsion. We remark that there exist alternative ways to study the soliton dynamics, for example, the variational approach used in Ref.~\cite{LiH}.

\begin{figure}[t]
\includegraphics[width=8.4cm]{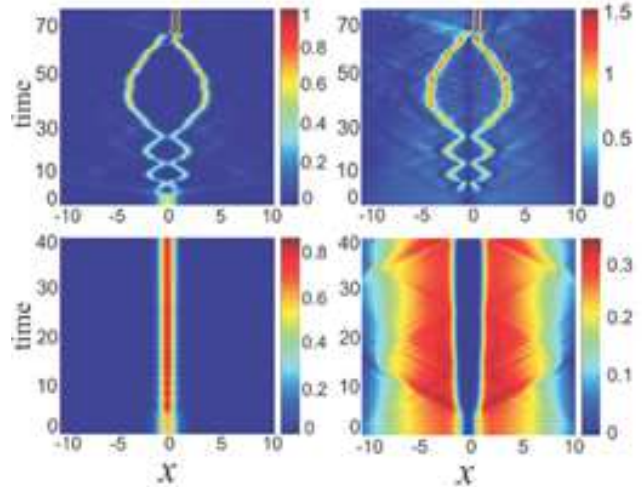}
\caption{(Color online) Conversion between different types of vector
solitons via Feshbach resonance.  The left and right plots represent
the longitudinal density profiles of the first and second species,
respectively. Units for $x$ and time are $\sqrt{\hbar/(m_1
\omega_{1\perp})}$ and $2\pi/\omega_{1\perp}$. Initially, a BD
vector soliton is prepared with $b_{12}=2$. Other parameters are the
same as in Fig.~\ref{BDDD}. Upper panel: From $t=3$ to 4, $b_{12}$
is linearly varied to a final value of $-10$ and stays at that value
afterwards. Lower panel: From $t=3$ to 23, $b_{12}$ is linearly
varied to a final value of 30 and stays at that value afterwards.}
\label{BD2BB}
\end{figure}

{\em Soliton transformation via Feshbach resonance {\rm --}}
Controlling the interaction strength via Feshbach resonace is a
unique possibility afforded by atomic systems. Fig.~\ref{BD2BB}
shows an example of dynamical conversion between different types of
vector solitons via an inter-species Feshbach resonance. Initially,
we prepare a zero-velocity BD vector soliton by choosing a proper
$b_{12}$ whose value is tuned in later time. In the upper panel of
Fig. \ref{BD2BB}, $b_{12}$ is changed to a value less than $\kappa
b_{11}$. For this new value of $b_{12}$, the system supports BB, but
not BD, vector solitons (see Fig.~\ref{demo-region}). The upper
panel of Fig.~\ref{BD2BB} shows that the initial BD vector soliton
is indeed destroyed after $b_{12}$ is changed to the new value. The
system first evolves into two pairs of BB vector-solitons with
finite relative velocity. These two pairs go through several
collisions and eventually merge into one single pair. During the
evolution, significant radiation from the second species is
observed. According to the exact BB solution given above, the number
of atoms in each species satisfy
\[ \frac{N_1}{N_2}=\frac{C_1}{-C_2}=\frac{b_{22}-\kappa
b_{12}}{\kappa b_{11} -b_{12}} \,.\]  Using realistic values of
$b_{11}$ and $b_{22}$ (as detailed in the caption of Fig. 1), we
have $N_1/N_2=0.935$. Numerically, we found this ratio to be $0.92
\sim 0.95$ which is in good agreement with the theoretical value. It
is simple to check and intuitive to understand that when the initial
BD vector-soliton breaks into two pairs of BB solitons, the relative
phase of the two bright solitons of the first species is zero, while
that of the second species is $\pi$. Just like in the scalar
condensate case, this leads to attractive (repulsive)
soliton-soliton interaction within the first (second) species. We
also remark that the relative velocity of the BB vector-soliton
pairs thus generated is sensitive to how fast $b_{12}$ is varied. In
general, the faster $b_{12}$ is varied, the larger the relative
velocity. If we vary $b_{12}$ much slower, the pairs (with smaller
relative velocity) will merge together later and experience more
collisions.  This observation suggests that varying the sweeping
rate of $b_{12}$  should help understand more about the collision
dynamics of multiple BB vector-soliton pairs ~\cite{pra, last}.

In the example shown in the lower panel of Fig.~\ref{BD2BB},
$b_{12}$ is changed to a value larger than $b_{22}/\kappa$, i.e.,
into the DD vector-soliton regime. Here the BD
vector soliton does not evolve into a DD
vector soliton, which is consistent with the fact that the DD
solution is unstable. Instead, due to the strong inter-species
repulsion, the core region of the initial dark soliton of species 2
is widened, and is no longer accurately described by a tanh
function. In the mean time, the bright soliton of species 1 is
narrowed and remain captivated inside the core of the dark
component. This property may help to produce and stabilize bright
solitons. During the evolution, we can also see that additional dark
solitons can be generated from the core region of species 2 (see the
lower right figure in Fig.~\ref{BD2BB}). Of particular interest are
those new dark solitons generated at early times: They move away from the
core, then get bounced back by the trapping potential, and finally
get reflected by the central BD vector soliton. Recently
similar behavior --- the reflection of dark soliton by a BD
vector soliton
--- was observed experimentally \cite{nature}.

\begin{figure}[h]
\includegraphics[width=8.5cm]{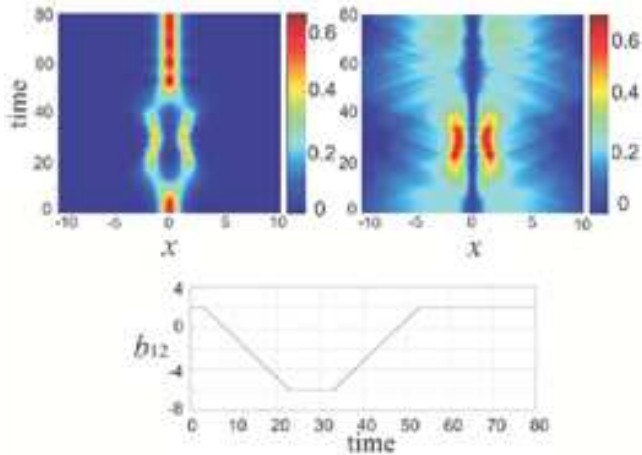}
\caption{(Color online) Conversion between BB and BD vector
solitons. The left and right plots of the upper panel represent the
longitudinal density profiles of the first and second species,
respectively. The lower panel shows the variation of $b_{12}$. Unit
for $x$ and $b_{12}$ is $\sqrt{\hbar/(m_1 \omega_{1\perp})}$, and
unit for time is $2\pi/\omega_{1\perp}$. Initially, a BD vector
soliton is prepared with $b_{12}=2$. Other parameters are the same
as in Fig.~\ref{BDDD}.} \label{BDBBBD}
\end{figure}

Figure ~\ref{BDBBBD} illustrates another example of the conversion between different
types of vector-solitons via inter-species Feshbach resonance. The initial
condition is the same as in Fig.~\ref{BD2BB} where a zero-velocity
BD vector-soliton is prepared. The value of $b_{12}$ is then
decreased and the system evolves to two BB vector-soliton pairs as
in Fig.~\ref{BD2BB}. Then, we restore the initial value of $b_{12}$
before the two BB vector-solitons merge into one. As seen in
Fig.~\ref{BDBBBD}, such a controlled tuning of $b_{12}$ can also
recover the initial BD vector-soliton. This represents a remarkable
example of the coherent control over the BEC dynamics, afforded by the
tunability of interaction strength.

In conclusion, we put forward a unified theory to treat all types of
vector solitons on equal footing in two-species BECs with arbitrary
scattering length. More importantly, we show that different types of
vector solitons can be transformed to each other by varying the
interaction strength.  Besides its fundamental interest in quantum
dynamics, our study may also find applications in the quantum
control of multi-species BECs. For example, treating one atomic
species as the target, its soliton properties can be controlled by
the presence of a second species via the tuning of the interaction
strength between them. We hope that these results can stimulate
investigation of novel phenomena in nonlinear complex systems in
general and further studies of BEC solitons in particular.

XXL was partially supported by the International Collaboration Fund,
National Univ. of Singapore. JG is supported by the start-up fund
(WBS grant No. R-144-050-193-101/133) and the ``YIA'' fund (WBS
grant No. R-144-000-195-123) from the National Univ. of Singapore.
HP acknowledges support from the NSF of US and the Welch Foundation
(Grant No. C-1669). WML is supported by the NSF of China under Grant
Nos. 10874235, 60525417, 10740420252, the NKBRSF of China under
Grant 2005CB724508, 2006CB921400.

\end{document}